\pdfoutput=1
\documentclass[10pt,twocolumn]{article}
\usepackage{styles/ieee-style}

\author{Mathias Gottschlag, Frank Bellosa}
\affiliation{Operating Systems Group\\ Karlsruhe Institute of Technology}
\email{os@itec.kit.edu}

\usepackage{styles/ka-style}
\usepackage{cite,xspace,ifthen,graphicx,listings}
\usepackage{styles/ka-style}

\usepackage[hyphens]{url}
\usepackage{color}
\usepackage{graphicx}
\usepackage{pgfplots}
\usepackage{comment}
\usepackage{csquotes}

\pgfplotsset{compat = 1.14}
\usetikzlibrary{arrows.meta}
\usetikzlibrary{calc}
\usetikzlibrary{patterns}
\usetikzlibrary{fit}

\usepackage[
   pdfauthor={Mathias Gottschlag, Frank Bellosa},
   pdftitle={Mechanism to Mitigate AVX-Induced Frequency Reduction},
   pdfkeywords={accelerators, performance isolation, frequency scaling, core specialization, AVX-512}
]{hyperref}

\definecolor{kitgreenexcl}{cmyk}{1.0,  0.0,  0.6, 0.0}
\definecolor{kitblue}     {cmyk}{0.8,  0.5,  0.0, 0.0}
\definecolor{kitgreen}    {cmyk}{0.6,  0.0,  1.0, 0.0}
\definecolor{kityellow}   {cmyk}{0.0,  0.05, 1.0, 0.0}
\definecolor{kitorange}   {cmyk}{0.0,  0.45, 1.0, 0.0}
\definecolor{kitbrown}    {cmyk}{0.35, 0.5,  1.0, 0.0}
\definecolor{kitred}      {cmyk}{0.25, 1.0,  1.0, 0.0}
\definecolor{kitpurple}   {cmyk}{0.25, 1.0,  0.0, 0.0}
\definecolor{kitcyan}     {cmyk}{0.9,  0.05, 0.0, 0.0}

\tikzset{
	hatch distance/.store in=\hatchdistance,
	hatch distance=10pt,
	hatch thickness/.store in=\hatchthickness,
	hatch thickness=2pt
}

\makeatletter
\pgfdeclarepatternformonly[\hatchdistance,\hatchthickness]{flexible hatch}
{\pgfqpoint{0pt}{0pt}}
{\pgfqpoint{\hatchdistance}{\hatchdistance}}
{\pgfpoint{\hatchdistance-1pt}{\hatchdistance-1pt}}%
{
	\pgfsetcolor{\tikz@pattern@color}
	\pgfsetlinewidth{\hatchthickness}
	\pgfpathmoveto{\pgfqpoint{0pt}{0pt}}
	\pgfpathlineto{\pgfqpoint{\hatchdistance}{\hatchdistance}}
	\pgfusepath{stroke}
}
\makeatother

\begin{document}

\title{Technical Report:\\Mechanism to Mitigate AVX-Induced Frequency Reduction}

\newcommand{\todo}[1]{{\texttt{[#1]}}}
\newcommand{\code}[1]{{\tt \small{#1}}}

\maketitle
\draftfooter

\begin{abstract}
	Modern Intel CPUs reduce their frequency when executing wide vector operations (AVX2 and AVX-512 instructions), as these instructions increase power consumption.
	The frequency is only increased again two milliseconds after the last code section containing such instructions has been executed in order to prevent excessive numbers of frequency changes.
	Due to this delay, intermittent use of wide vector operations can slow down the rest of the system significantly.
	For example, previous work has shown the performance of web servers to be reduced by up to 10\% if the SSL library uses AVX-512 vector instructions\cite{krasnovdangers}.
	These performance variations are hard to predict during software development as the performance impact of vectorization depends on the specific workload.

	We describe a mechanism to reduce the slowdown caused by wide vector instructions without requiring extensive changes to existing software.
	Our design allows the developer to mark problematic AVX code regions.
	The scheduler then restricts execution of this code to a subset of the cores so that only these cores' frequency is affected.
	Threads are automatically migrated to a suitable core whenever necessary.
	We identify a suitable load balancing policy to ensure good utilization of all available cores.
	Our approach is able to reduce the performance variability caused by AVX2 and AVX-512 instructions by over 70\%.
\end{abstract}

\section{Introduction}

Parallelization of computation-intensive code with SIMD instructions can cause significant performance improvements.
In particular, wider vector instructions can often improve throughput.
For example, the OpenSSL implementation of the ChaCha20-Poly1305 encryption algorithm achieves a throughput of up to 2.89\,GB/s, whereas the implementation in the BoringSSL library only achieves 1.6\,GB/s\cite{krasnovdangers}.
One of the reasons for this difference is that BoringSSL only uses 256-bit wide AVX2 operations whereas OpenSSL employs 512-bit wide AVX-512 instructions.

The end of Dennard Scaling has lead to a situation where modern CPUs cannot utilize the full silicon area at maximum frequency anymore\cite{taylor2012dark}.
Instead, CPUs either have to keep parts of the chip unused or have to reduce their operating frequency.
As chip area utilization depends on the complexity of the instructions executed -- resulting in significantly varying power consumption\cite{molka2010characterizing} -- it is possible to increase CPU frequency when only instructions of lower complexity are executed.
In particular, the parallel operations of wide SIMD instructions can temporarily draw large amounts of additional current.
Therefore, recent Intel CPUs limit their frequency depending on the rate and type of executed SIMD instructions.
For example, the Intel Xeon Silver 4116 reduces its base frequency from 2.1\,GHz to 1.1\,GHz when executing certain AVX-512 instructions\cite{xeonscalableerrata}.

These frequency changes are associated with some overhead, though\cite{mazouz2014evaluation}.
Current Intel server CPUs only revert frequency changes caused by AVX instructions approximately 2\,ms after the last problematic section of code has been executed\cite[Section 15.26]{optimizationmanual}, likely in order to limit the rate of frequency changes.
This delay can have a significant negative impact on the performance of systems where only parts of the workload are vectorized, as a large amount of scalar (i.e., non-vectorized) code following the problematic AVX code is slowed down as well.
For example, in web server benchmarks conducted at Cloudflare\cite{krasnovdangers} to compare the two SSL libraries mentioned above, a system using OpenSSL performed 10\% worse overall than a system using BoringSSL, even though the SSL library in isolation was almost twice as fast.

This example shows that vectorized code can have a negative impact on completely unrelated sections of code, which is highly problematic for several reasons:
\begin{itemize}
	\item \textbf{Breach of isolation:}
		If one process executes vectorized code which reduces the CPU frequency, all other processes which are scheduled in the next two milliseconds also execute with reduced CPU frequency.
		This effect not only poses a problem for scheduling fairness, as subsequent processes suffer from reduced performance, but also allows the construction of a covert channel between two otherwise isolated processes where information is transmitted via frequency changes.
	\item \textbf{Optimization complexity:} Traditionally, software developers expect a high degree of performance composability, i.e., optimizations in one software component do not negatively affect the performance of other components.
		If vectorization of one component reduces the CPU frequency for other components, this assumption is broken and all optimization work on a single component has to take the whole software system into account.

		Furthermore, the impact of frequency changes on overall performance depends on the amount of scalar code in between vectorized parts, so vectorization also has to take all potential workloads into account.
		In the case of software libraries such as OpenSSL, it is hard to predict in which environment they are going to be used.
	\item \textbf{Performance predictability:}
		As it is hard to predict performance implications of vectorization, inconspicuous updates of software components promising performance improvements can to the contrary cause significant performance problems (as, for example, observed at Cloudflare\cite{krasnovdangers}).
		In the worst case, vectorized code can negatively affect the quality of the service provided by the system or can cause the system to fail to meet realtime requirements, posing a threat to system reliability.
\end{itemize}

Although these performance variability problems are currently limited to recent generations of CPUs and the use of AVX2 and AVX-512, we expect future CPUs to show similar behaviour.
The underlying mechanism -- increasing the CPU frequency when a limited subset of the supported instructions is used -- is a sensible optimization if power consumption varies significantly depending on the executed instructions.
If the amount of chip area used for accelerators increases, the variation of power consumption increases as well, which in turn increases the scale of the resulting frequency differences.
As the performance of CPUs is increasingly limited by their power consumption, literature suggests an increasing amount of task-specific accelerators as one method to increase the power efficiency and therefore the performance of CPUs\cite{taylor2012dark}.

Previous work to optimize existing software with AVX-512 has lead Tiwari et al. to the suggestion that in the future the software should be rearchitected to split parts using AVX-512 into different threads and concentrate those threads on a subset of the system's CPUs\cite{tiwari2018accelerating}.
Such a split would limit the impact of the frequency drop to specific cores.
However, such changes usually require significant engineering effort.

In this work, we propose thread migration as an unintrusive mechanism to limit execution of wide vector instructions to a subset of the system's cores in order to limit the frequency effects to these cores.
If the operating system automatically migrates threads to suitable cores whenever they start or stop executing AVX instructions, there is no need for any significant restructuring of the application.
We describe a scheduling policy which is able to limit AVX instructions to few cores and which employs load balancing and prioritizing to achieve full utilization of all cores.
We also describe a workflow which can be used to identify code sections which have the potential to cause significant frequency reduction, and provide a simple method to mark those code sections to trigger the appropriate thread migrations.

Even though thread migration has the potential to cause higher overhead than other application-specific mechanisms to isolate problematic vectorized code sections, we show that the overhead is low enough to reduce performance variation by over 70\% in a web server scenario similar to the one described above.

\section{Analysis}
\label{sec:analysis}

For recent Xeon Scalable processors, Intel documents three different sets of maximum frequencies due to the heat output and current requirements of different instructions.
For example, the Xeon Gold 6130 processor has an all-core turbo frequency of 2.8\,GHz for most instructions (\emph{frequency level 0} in Intel parlance), whereas the all-core turbo frequency for heavy AVX-512 instructions (multiplications and fused multiply-add) is documented to be 1.9\,GHz (\emph{frequency level 2})\cite{xeonscalableerrata}.
All other AVX-512 instructions as well as heavy AVX2 instructions allow for an intermediate all-core turbo of 2.4\,GHz (\emph{frequency level 1}).

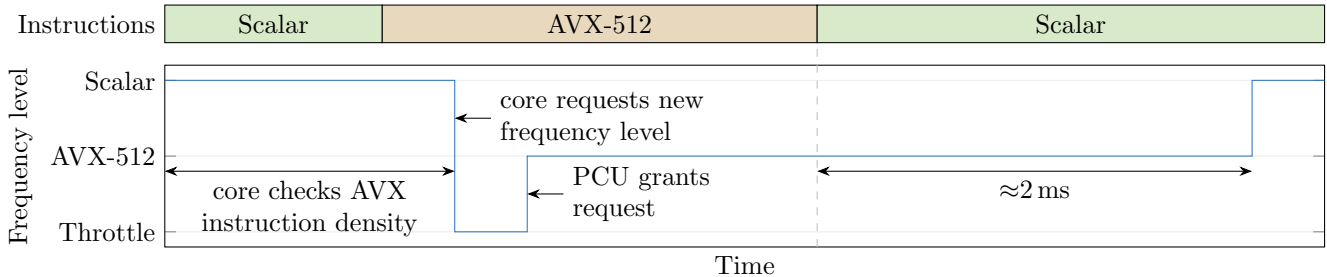
\begin{figure*}
	\begin{center}
	\begin{tikzpicture}
		\begin{axis}[
			legend pos=north east,
			ylabel={Frequency level},
			xlabel={Time},
			ytick={0, 1, 2},
			yticklabels = {Throttle,AVX-512,Scalar},
			xmajorticks=false,
			width=17cm,
			height=4cm,
			grid style={line width=.1pt, draw=gray!15},
			ymajorgrids,
			xmin=0,
			xmax=16,
		]
		\addplot[color=kitblue] coordinates {
			(0,2)
			(4,2)
			(4,0)
			(5,0)
			(5,1)
			(15,1)
			(15,2)
			(16,2)
		};
		\draw[black, thin, {Stealth}-{Stealth}] (9,0.8) -- node[below] {$\approx$2\,ms} (15,0.8);
		\draw[black, thin, {Stealth}-{Stealth}] (0,0.8) -- node[below, align=center] {core checks AVX\\instruction density} (4,0.8);

		\draw[black, thin, {Stealth}-{}] (4,1.5) -- (4.5,1.5) node[right, align=left] {core requests new\\frequency level};
		\draw[black, thin, {Stealth}-{}] (5,0.5) -- (5.5,0.5) node[right, align=left] {PCU grants\\request};

		\coordinate (topleft) at (current axis.north west);
		\coordinate (topright) at (current axis.north east);
		\coordinate (avxstart) at ($(topleft) + (3,0)$);
		\coordinate (avxend) at ($(topleft) + (9,0)$);
		\coordinate (avxendbottom) at ($(current axis.south west) + (9,0)$);
		\coordinate (avxstartbottom) at ($(current axis.south west) + (3,0)$);
		\end{axis}
		\draw [fill=kitgreen!25] ($(topleft) + (0, 0.3)$) rectangle ($(avxstart) + (0, 0.8)$) node[pos=.5] {Scalar};
		\draw [fill=kitbrown!25] ($(avxstart) + (0, 0.3)$) rectangle ($(avxend) + (0, 0.8)$) node[pos=.5] {AVX-512};
		\draw [fill=kitgreen!25] ($(avxend) + (0, 0.3)$) rectangle ($(topright) + (0, 0.8)$) node[pos=.5] {Scalar};
		\draw[black, thin] ($(topleft) + (0, 0.55)$) node[left, align=left] {Instructions};
		\draw [black!25, dashed] (avxendbottom) -- ($(avxend) + (0, 0.3)$);
	\end{tikzpicture}
	\end{center}

	\caption{
		Frequency levels when an Intel Skylake-SP core temporarily executes 512-bit FMA instructions.
		After AVX-512 usage has been detected, the core executes at reduced performance while requesting a new power license level\cite{optimizationmanual}.
		Once the request has been granted, the core switches to the new frequency.
	}
	\label{fig:freqswitch}
\end{figure*}

The frequency is, however, only reduced if at least approximately one instruction of the corresponding type is executed per cycle\cite{lemireavx} or if a sufficiently dense mixture of instructions from two different categories is executed\cite[Section 15.26]{optimizationmanual}.
This frequency reduction affects each core individually depending on the workload on the specific core\cite{optimizationmanual}.

The transition between different frequency levels does not happen instantaneously when the instruction mix changes.
Instead, the frequency reduction consists of multiple steps as shown in the left half of Figure~\ref{fig:freqswitch}.
First, the core needs to recognize that the condition for a frequency change has been fulfilled.
Then, the core requests a different \emph{power license} from the central \emph{package control unit} (PCU).
The PCU takes up to 500 microseconds to evaluate the number of cores executing at the new license level, during which the core operates with reduced performance\cite[Section 15.26]{optimizationmanual}.
Only afterwards can the core switch to the frequency level for the detected instruction mix.
Switching back to a higher frequency level is, as described above, delayed as well\cite{optimizationmanual}.
When the CPU detects that the conditions for the frequency reduction are not fulfilled anymore, it waits approximately two milliseconds before reverting the frequency change.

The latter mechanism in particular can negatively impact the performance of the scalar code following AVX instructions.
To analyze the extent of this effect, we replicate the web server benchmarks conducted at Cloudflare\cite{krasnovdangers}.
We let the nginx web server serve static files via HTTPS.
The server uses the ChaCha20-Poly1305 encryption algorithm from OpenSSL library, which we compile for different SIMD instruction sets.
To increase the scalar parts of the workload, the server optionally compresses the files on-the-fly with the brotli compression algorithm.

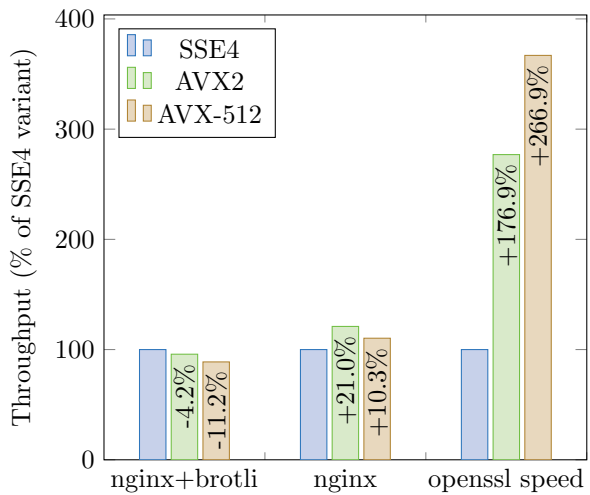
\begin{figure}[h]
	\begin{tikzpicture}
		\begin{axis}[
			ybar,
			legend pos=north west,
			ylabel={Throughput (\% of SSE4 variant)},
			symbolic x coords={
				Workload 1,
				Workload 3,
				Workload 4
			},
			enlarge x limits=0.25,
			xtick=data,
			xticklabels = {
				nginx+brotli,
				nginx,
				openssl speed
			},
			nodes near coords,
			every node near coord/.append style={color=black,rotate=90, anchor=east},
			ymin=0,
			width=8cm,
			height=7.5cm,
			point meta=explicit symbolic,
			xtick align=inside,
			major x tick style = {opacity=0},
			minor x tick num = 1,
		]
		\addplot[color=kitblue, fill=kitblue!25] coordinates {
			(Workload 1,100) []
			(Workload 3,100)  []
			(Workload 4,100)  []
		};
		\addplot[color=kitgreen, fill=kitgreen!25] coordinates {
			(Workload 1,95.8) [-4.2\%]
			(Workload 3,121.0)  [+21.0\%]
			(Workload 4,276.9)  [+176.9\%]
		};
		\addplot[color=kitbrown, fill=kitbrown!25] coordinates {
			(Workload 1,88.8) [-11.2\%]
			(Workload 3,110.3) [+10.3\%]
			(Workload 4,366.9)  [+266.9\%]
		};
		\legend{SSE4, AVX2, AVX-512}


		\end{axis}
	\end{tikzpicture}
	\caption{Different workloads show different sensitivity to the frequency reduction caused by wide SIMD instruction sets (normalized to SSE4 performance).}
	\label{fig:workloads}
\end{figure}

Figure~\ref{fig:workloads} shows the results of this test as well as the throughput of a microbenchmark exercising only the encryption algorithm.
It is immediately visible that, although AVX2- and AVX-512-enabled cryptography performs worse for a web server serving compressed websites, AVX2 has significant advantages if the web server serves uncompressed data, while in microbenchmarks the AVX-512 cryptography code provides the highest performance.
The choice of instruction sets therefore depends on the workload, yet software developers of generic libraries can rarely predict the user's workload and usually optimize for what they expect to be the most likely workload or only optimize the performance in microbenchmarks.

\subsection{Core Specialization for Scalar Code}

Software libraries could provide various implementations of each algorithm and could either automatically decide at runtime which SIMD instruction set to use or could let the user programmatically choose the specific implementation.
However, disabling vectorization can -- at least locally -- be detrimental to performance, as the unvectorized codepaths usually execute slower than their vectorized counterparts.
Also, most existing software libraries do not allow for the user to select the type of vectorization.
Therefore, a more promising approach is to leave the application code unchanged, but isolate the vectorized code in order to limit its negative effects on overall performance.

Whereas on Intel Haswell CPUs the frequency drop caused by AVX code always affects all cores in the system, starting from the Broadwell microarchitecture the CPUs only reduce the clock of individual cores executing AVX instructions\cite{optimizationmanual}.
Therefore, one method to isolate vectorized code sections is to place AVX code and scalar code sections on different sets of cores.
Assuming that a fine-grained partitioning of the program into predominantly scalar code sections and code sections containing significant amounts of AVX instructions is available, applying such \emph{core specialization} would limit the negative impact on frequency to those cores which only execute AVX code sections.

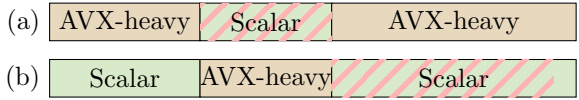
\begin{figure}
	\begin{center}
	\begin{tikzpicture}
		\draw [] (0, 1) node[left, align=left] {(a)};
		\draw [fill=kitbrown!25] (0, 0.75) rectangle (2, 1.25) node[pos=.5] {AVX-heavy};
		\draw [fill=kitgreen!25] (2, 0.75) rectangle (3.75, 1.25);
		\draw [fill=kitbrown!25] (3.75, 0.75) rectangle (7, 1.25) node[pos=.5] {AVX-heavy};
		\fill [pattern=flexible hatch, pattern color=red!30] (2, 0.75) rectangle (3.75, 1.25);
		\node[fit={(2,0.75) (3.75,1.25)}, inner sep=0pt, label=center:Scalar] (A) {};

		\draw [] (0, 0.25) node[left, align=left] {(b)};
		\draw [fill=kitgreen!25] (0, 0) rectangle (2, 0.5) node[pos=.5] {Scalar};
		\draw [fill=kitbrown!25] (2, 0) rectangle (3.75, 0.5) node[pos=.5] {AVX-heavy};
		\draw [fill=kitgreen!25] (3.75, 0) rectangle (7, 0.5);
		\fill [pattern=flexible hatch, pattern color=red!30] (3.75, 0) rectangle (6.7, 0.5);
		\node[fit={(3.75,0) (7,0.5)}, inner sep=0pt, label=center:Scalar] (A) {};
	\end{tikzpicture}
	\end{center}

	\caption{
		It is less problematic if a core mostly executes AVX-heavy code and intermittently executes predominantly scalar code (a) than vice versa (b), as the latter causes a frequency reduction for at least two milliseconds.
		The hatched parts are scalar code slowed down by AVX-heavy code.
	}
	\label{fig:asymmetrictaskstealing}
\end{figure}

A strict partitioning of cores into cores which only execute mostly scalar code (in the following called \emph{scalar cores}) and cores which only execute AVX-heavy code (\emph{AVX cores}), however, results in reduced performance because, unless the ratio of cores perfectly matches the ratio of time spent in scalar code versus AVX code, one of the two sets of cores will necessarily be underutilized.
Choosing the correct ratio of cores will lead to a situation with high utilization where the idle time of cores between two code sections of the appropriate type is low.
Even then, however, significant amounts of idle time exist, which significantly reduces the overall system performance.
Achieving full CPU utilization is only possible if either AVX cores temporarily execute scalar code when they would be otherwise idle, or vice versa.

As the time periods where no suitable code is available are rather short, the effect of the executed code on CPU frequency has to be considered.
Figure~\ref{fig:asymmetrictaskstealing} shows the periods of reduced CPU frequency when a AVX core intermittently executes scalar code (a) and when scalar code executes AVX-heavy code (b).
The figure shows that the former case is less problematic, as only the short section of scalar code is unnecessarily slowed down, whereas in the latter case every short section of AVX code slows down two milliseconds worth of scalar code.
This asymmetry has to be taken into account during allocation of cores for AVX code (the scheduler must allocate enough cores) and during scheduling (scalar cores must never execute AVX-heavy code).

\section{Design and Implementation}

As described above, limiting the use of wide vector operations to specific cores can significantly improve performance.
Most applications, however, do not use a threading model which allows individual operations to be executed on a different thread on a different core.
Reengineering these applications to introduce a separate thread pool for code sections with wide vector operations is likely excessively expensive and time-consuming.

\begin{figure}
	\begin{center}
		\begin{tabular}{c}{
			\begin{lstlisting}[frame=single,linewidth=0.4\textwidth]
/* [...] */
with_avx();
n = SSL_read(c->ssl->connection, buf, size);
without_avx();
/* [...] */
			\end{lstlisting}
		}\end{tabular}
	\end{center}
	\caption{
		Example of an annotated function call from nginx which potentially executes AVX-512 instructions.
		The call to \emph{with\_avx()} marks the task as an AVX task and migrates the task to an AVX core, whereas the call to \emph{without\_avx()} reverts the task type change and potentially migrates the task to a scalar core.
	}
	\label{fig:api}
\end{figure}

Therefore, we suggest migration of mostly unmodified threads between cores as a mechanism to isolate scalar code from AVX code, so that the system will automatically migrate threads to the appropriate core.
Combined with fault-and-migrate to recognize AVX code sections\cite{li2010operating}, such a mechanism can be used to create a fully-automatic solution for existing applications.
In our prototype, instead, the application developer manually marks sections of code if they potentially execute expensive vector instructions.
Although such an approach requires some manual intervention, no large changes to the application are required.
Figure~\ref{fig:api} shows an example for the annotation mechanism used by our design.
Before and after the function which potentially executes AVX-512 instructions, we have placed a system call which informs the OS about whether the thread is going to execute wide vector instructions in the future or not.
In the following, we name threads either \emph{AVX tasks} or \emph{scalar tasks} depending on whether they expect to execute wide vector instructions or not.

\subsection{Core Specialization}

As described above, it is more problematic to allow an AVX task to execute on a core which usually only executes scalar tasks than to let a core which mostly executes AVX tasks to execute a scalar task.
Therefore, our approach limits the set of available cores for AVX tasks (in the following called \emph{AVX cores}).
Whereas AVX cores can execute scalar tasks if necessary, all other cores must only execute scalar tasks.

The system therefore allocates as many AVX cores as required for the AVX tasks in the system or more and those cores prioritize AVX tasks, but execute arbitrary scalar tasks if no AVX task is available.
If a thread becomes an AVX task while it is still running on a scalar core, the scheduler immediately suspends the thread and schedules a scalar task instead.
The (now suspended) AVX task preempts any scalar task running on an AVX core.
As core specialization is implemented as a restriction of the allowed cores for AVX tasks, the regular scheduler load balancing mechanism automatically migrates tasks to the appropriate cores.

\subsection{Scheduler Implementation}

We implement the policy described in the last section as an extension of the MuQSS scheduler\cite{muqss}, an alternative out-of-tree scheduler for Linux.
We chose the MuQSS scheduler over the CFS scheduler due to its significantly lower code complexity: Whereas in Linux 4.17 MuQSS consists of 7326 lines of code, much of which is common boilerplate code also found in the CFS scheduler, CFS encompasses 23399 lines and implements a more complex policy, making an extension substantially harder\footnote{We count the lines in all files which are only compiled for one of the two schedulers and which cannot be disabled by a separate configuration option.}.
While our implementation modifies some of the core data structures of MuQSS, we expect an implementation in a different scheduler to be viable.

The central data structure of MuQSS is a run queue sorted by the tasks' virtual deadline\cite{muqss}.
Depending on the configuration, MuQSS allocates one such run queue per system, per logical CPU or -- as chosen for our experiments, as it maximizes throughput -- per physical core.
We replicate each run queue of MuQSS three times in order to separate the different types of tasks.
The original MuQSS code always selects the task with the earliest deadline.
This replication of run queues allows us to restrict the types of tasks executed on a core or to deprioritize certain types of tasks by adding a penalty to the deadline of all tasks in the corresponding run queues.
The first two run queues contain the scalar tasks and AVX tasks respectively.
The last run queue contains all tasks which have never declared the type of their work (i.e., all tasks not belonging to the instrumented application), as these tasks are not restricted to run on any specific core.
In particular, the last run queue contains system tasks pinned to AVX cores.
If these tasks were treated like scalar tasks, they would be starved by AVX tasks.

A scalar core only picks tasks from the first and the last run queue in order to prevent the execution of AVX tasks.
An AVX core, instead, can pick tasks from all run queues, but only runs scalar tasks if no other runnable tasks are available to the core.
This priority scheme is implemented by adding a large value to the deadline of scalar tasks so that the deadline of all other tasks is guaranteed to be lower.
Note that a similar mechanism is already used in MuQSS to implement tasks with idle priority.

In the MuQSS scheduler, whenever a core selects the next task, it also (locklessly) checks the minimum deadline of the run queues of all other cores and steals a task from a different core if that task has a lower deadline.
This mechanism is responsible for all load balancing between scalar and AVX cores in our prototype.
Whenever on a scalar core a scalar task becomes an AVX task, it is put back into the local run queue and if there is any scalar task executing on an AVX core, that task is preempted via an inter-processor interrupt to allow the AVX core to select the new AVX task instead.

\subsection{Identifying Problematic Code}

To implement core specialization, the scheduler requires information about whether tasks are scalar or AVX tasks.
As shown in Figure~\ref{fig:api}, this information is provided by the developer in form of manual instrumentation of AVX code sections.
As analyzing complex code bases in order to identify vectorized functions can be time consuming, we provide tools and a corresponding workflow to help the developer to identify problematic code regions.

First, a static analysis tool disassembles the target application as well as all its dynamically linked libraries and analyzes the usage of wide vector registers.
For every function, the program calculates the ratio between the number of the instructions accessing 256-bit and 512-bit registers and the total instruction count.
As the frequency reduction depends on the frequency of wide vector operations\cite{lemireavx}, this ratio is a good indicator for whether the function is causing a frequency drop and for the magnitude of the frequency drop.
As a result, the program prints a list of functions sorted by this AVX instruction ratio.
The functions with a high ratio are good candidates for core specialization.

This static analysis limits the number of functions which need to be considered.
However, whether a function causes a significant frequency drop depends on a number of additional factors, and frequently called functions such as memcpy should not cause the thread to migrate to a different core if they do not cause any frequency change.
Intel documents some conditions for frequency changes\cite{optimizationmanual}.
Whether these conditions are fulfilled largely depends on the microarchitecture.
For example, a sufficiently dense mix of AVX-512 and AVX2 instructions causes a switch to the AVX-512 frequencies, but pipeline stalls during execution due to dependencies can cause the vector instruction frequency to be decreased enough to prevent frequency changes.

In some cases the output of the static analysis clearly identifies the problematic code.
For the remaining cases, as accurate modelling of the microarchitecture is impractical, we suggest the use of performance counters to determine whether a function causes frequency changes.
To track the frequency level, Intel provides four performance counters which count the cycles spent at each frequency level.
The \emph{CORE\_POWER.LVL0\_TURBO\_LICENSE} performance event as well as the \emph{LVL1\_TURBO\_LICENSE} and \emph{LVL2\_TURBO\_LICENSE} events count the cycles spent at the frequency levels for scalar code, AVX2 code and AVX-512 code as described in Section~\ref{sec:analysis}, respectively, whereas the CORE\_POWER.THROTTLE event counts the cycles with reduced performance during a power license request\cite{optimizationmanual}.
The core requests a new power license whenever the condition for a frequency reduction has been fulfilled, i.e., when the core has determined that it exceeds the current power budget.

As shown in Figure~\ref{fig:freqswitch}, both entering and exiting a lower frequency level is delayed, which restricts the performance counters which can be used to detect problematic code.
In particular, reverting to the frequency for scalar code is delayed by at least two milliseconds, so the counters for the AVX2 and AVX-512 frequency levels cannot be reliably used to determine the code which caused the frequency drop, as they are incremented during all following scalar code affected by the frequency reduction.
Also, during the transition, the CPU throttles not just during the offending AVX code but also for some time afterwards while waiting for the PCU to grant the new power license.
However, the time spent waiting is significantly shorter (up to 0.5\,ms) than the time spent at the AVX frequency levels, and throttling begins right after the condition for the frequency reduction has been detected.

The latter property in particular is why the CORE\_POWER.THROTTLE counter event is a good indicator for code which causes frequency changes.
After completing the static analysis, the user therefore generates a flame graph\cite{gregg2016flame} from this counter.
A flame graph is a tool to visualize where in the call tree of an application a performance counter is increased.
In most cases, the flame graph is used to visualize CPU cycles to locate hot spots in the application.
Visualizing THROTTLE cycles instead of all CPU cycles shows approximately where in the call tree frequency changes are triggered.

Because, as described above, frequency changes are delayed by up to 0.5\,ms and therefore unrelated code might be shown in the flame graph, the functions shown in the graph have to be compared to the output of the static analysis pass to remove false positives.
If core specialization is not able to completely eliminate frequency changes on the scalar cores because not all problematic code segments have been identified, the developer can repeat the performance counter analysis to identify the remaining AVX code.

Note that, as described in \ref{sec:analysis}, there is a short delay of up to approximately 100 instructions between the execution of the first instruction of an AVX-heavy code section and the time at which the core recognizes the need for a lower frequency level.
This delay might cause the performance counter analysis to miss very short AVX-heavy functions.
As AVX instructions are particularly useful to parallelize operations on data streams, we expect such short AVX instruction bursts to be highly unusual.
In addition, due to the conditions for frequency changes\cite{optimizationmanual,lemireavx}, we expect such short AVX instructions to have little effect on CPU frequencies.

It is, however, possible to detect even such code sections if they cause frequency changes, even though the execution of these code sections already lies in the past at the moment the CPU recognizes the need for a frequency change.
If the operating system configures the performance counter to overflow at the very first CORE\_POWER.THROTTLE cycle, the counter overflow interrupt can be used to notify the OS.
The OS can then examine the \emph{last branch records}\cite{kleen2014lbr} of the CPU to determine the recently executed code which caused the frequency change.
We did not find any workload which had sufficiently short AVX code sections, therefore we did not implement this technique.

\section{Evaluation}

To show that our approach is able to reduce the performance variation caused by vectorized parts of a program, we evaluate our design in a scenario similar to the experiments conducted at Cloudflare\cite{krasnovdangers}.
We run the nginx web server and let it serve a compressed static page via HTTPS.
The connections are encrypted with the ChaCha20-Poly1305 encryption algorithm, with the implementation provided by the OpenSSL library.
For the different experiments, OpenSSL is compiled with support for either AVX-512, AVX2 or only SSE4 vector instructions.

All experiments are run on a system with an Intel Xeon Gold 6130 processor and 24\,GiB of DDR4 RAM.
The processor contains 16 physical cores with all-core turbo frequencies of 1.9\,GHz, 2.4\,GHz and 2.8\,GHz for heavy AVX-512 instructions, heavy AVX2 instructions and other instructions respectively.
The web server is executed on 12 of the 16 cores, and the wrk2 web server benchmark client is executed on the other 4 cores to generate HTTPS requests.

Static analysis showed use of AVX2 and AVX-512 instructions in the OpenSSL implementation of ChaCha20 and Poly1305, in one function in glibc's profiling code, and in memset/memcpy/memmove.
Analysis of the CORE\_POWER.THROTTLE performance counter showed that only OpenSSL encryption and decryption code caused frequency changes.
Therefore, we annotated the calls to the OpenSSL functions \emph{SSL\_read, SSL\_write, SSL\_do\_handshake} and \emph{SSL\_shutdown} to restrict execution of these functions to the last two physical cores.
In total, annotations added only 9 lines to the program.

\begin{figure}
	\begin{tikzpicture}
		\begin{axis}[
			ybar,
			legend pos=south west,
			ylabel={Throughput ($\times$1000\,req/s)},
			symbolic x coords={SSE4, AVX2, AVX-512},
			enlarge x limits=0.25,
			xtick=data,
			nodes near coords,
			every node near coord/.append style={color=black,rotate=90, anchor=east},
			ymin=0,
			width=8cm,
			height=6cm,
			point meta=explicit symbolic,
			major x tick style = {opacity=0},
			minor x tick num = 1,
			xtick align=inside,
		]
		\addplot[color=kitblue, fill=kitblue!25, error bars/.cd, y dir=both, y explicit] coordinates {
			(SSE4, 6.882) += (0,0.011) -= (0,0.011) [] 
			(AVX2, 6.593) += (0,0.012) -= (0,0.012) [-4.2\%] 
			(AVX-512, 6.110) += (0,0.014) -= (0,0.014) [-11.2\%] };
		\addplot[color=kitgreen, fill=kitgreen!25, error bars/.cd, y dir=both, y explicit] coordinates {
			(SSE4, 6.876) += (0,0.024) -= (0,0.024) [-0.1\%] 
			(AVX2, 6.804) += (0,0.006) -= (0,0.006) [-1.1\%] 
			(AVX-512, 6.660) += (0,0.007) -= (0,0.007) [-3.2\%] };
		\legend{Unmodified, Core Specialization}
		\end{axis}
	\end{tikzpicture}
	\caption{
		Throughput of nginx with OpenSSL compiled for different instruction sets:
		The blue bars show the throughput of the unmodified web server, whereas the green bars indicate performance when execution of SSL code is restricted to a subset of the system's cores.
	}
	\label{fig:throughput}
\end{figure}
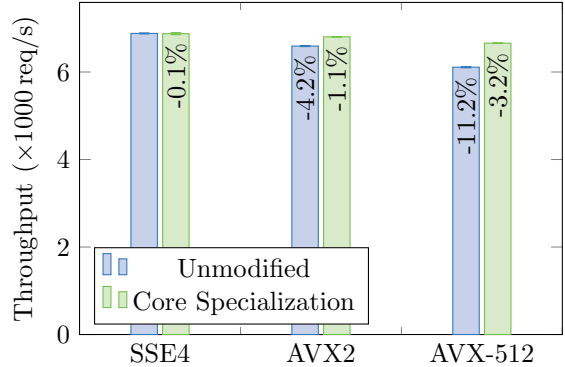

Figure~\ref{fig:throughput} shows the throughput of the web server in the scenario described above with and without core specialization.
Without core specialization, throughput varies depending on the SIMD instruction set used in the SSL library.
For AVX2 and AVX-512 instructions, throughput is reduced by 4.2\% and 11.2\%, respectively, over a version using only instructions which do not cause any frequency drop.
With core specialization, this negative effect of vectorization on performance is significantly reduced:
The performance drop is reduced to 1.1\% for AVX2 and to 3.2\% for AVX-512 instructions, a reduction by 74\% and 71\% respectively.
These results show that core specialization is able to significantly reduce performance variability and causes little overhead, even though additional scheduler invocations are performed.

\subsection{CPU Frequency}

\begin{figure}
	\begin{tikzpicture}
		\begin{axis}[
			ybar,
			legend pos=south west,
			ylabel={Frequency (GHz)},
			symbolic x coords={SSE4, AVX2, AVX-512},
			enlarge x limits=0.25,
			xtick=data,
			nodes near coords,
			every node near coord/.append style={color=black,rotate=90, anchor=east},
			ymin=0,
			width=8cm,
			height=6cm,
			point meta=explicit symbolic,
			major x tick style = {opacity=0},
			minor x tick num = 1,
			xtick align=inside,
		]
		\addplot[color=kitblue, fill=kitblue!25, error bars/.cd, y dir=both, y explicit] coordinates {
			(SSE4, 2.800) += (0,0.000) -= (0,0.000) [] 
			(AVX2, 2.676) += (0,0.001) -= (0,0.001) [-4.4\%] 
			(AVX-512, 2.481) += (0,0.001) -= (0,0.001) [-11.4\%] };
		\addplot[color=kitgreen, fill=kitgreen!25, error bars/.cd, y dir=both, y explicit] coordinates {
			(SSE4, 2.791) += (0,0.013) -= (0,0.013) [-0.3\%] 
			(AVX2, 2.749) += (0,0.001) -= (0,0.001) [-1.8\%] 
			(AVX-512, 2.687) += (0,0.002) -= (0,0.002) [-4.0\%] };
		\legend{Unmodified, Core Specialization}
		\end{axis}
	\end{tikzpicture}
	\caption{
		Average frequency of the cores executing the nginx web server.
		Core specialization limits the frequency reduction to 2 out of 12 cores.
		Note that the frequency improvement achieved by core specialization is lower because the unmodified web server is partially able to run at maximum frequencies.}
	\label{fig:frequency}
\end{figure}
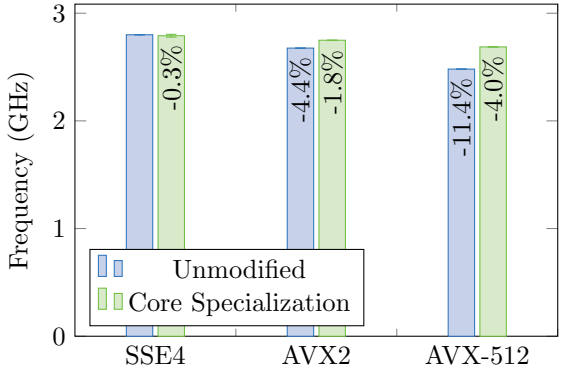

The main reason for the performance improvement lies in the improved CPU frequencies, as the goal of core specialization is to limit the AVX-induced frequency drop to a subset of the cores.
In our benchmark scenario, only two out of 12 cores should be affected by AVX code.
Note, however, that this does not automatically result in a six-fold reduction of any frequency drops.
As the original software is partially able to run the cores at maximum frequency whereas core specialization causes a significantly higher concentration of AVX code on the AVX cores, we expect a lower effect on CPU frequencies, which is mirrored by the performance measurements shown above.

To show the effect on frequency, we repeat the experiment from the last section and measure the average frequency across all cores.
The results are shown in Figure~\ref{fig:frequency}.
For AVX2 code, the frequency drop is reduced from 4.4\% to 1.8\%, and for AVX-512 code from 11.4\% to 4.0\%.
These numbers closely correlate with the measured performance.

As performance is affected by other factors such as overhead as well as different cache utilization, the results slightly differ, though.
Therefore, in the next sections, we analyze these factors to give a better overview over the effects of core specialization on performance.

\subsection{Instructions per Cycle}

One particularly striking result is that our prototype provides higher performance than expected from the CPU frequency measurements alone, even though thread migration and frequent scheduler invocations should on the contrary cause additional overhead.
As expected, measurements conducted with nginx when using SSE4 OpenSSL show a slightly increased number of instructions executed per HTTPS request (+0.7\%) when using core specialization.
Performance counter analysis of the instructions per cycle, however, also shows an improvement (+0.7\%).
The improved utilization of the CPU therefore makes up for the overhead caused by the scheduler invocations.

To determine the cause of the IPC improvement, we conduct a performance counter analysis with Intel VTune Amplifier.
The analysis shows that the system with core specialization experiences slightly more stall cycles due to memory accesses.
However, core specialization also significantly reduces the number of mispredicted branches due to more effective use of the misprediction tables.
These tables exist once per core and cache branch history, and restricting the amount of code executed on a core reduces the amount of code that needs to be covered by the respective branch predictor.
Similar effects on caches have already been shown for other approaches using core specialization or cohort scheduling\cite{larus2002using,kallurkar2017schedtask,harizopoulos2004steps}.

\subsection{Overhead of Thread Migration}

As shown in the previous experiment, it is difficult to measure the overhead of frequent context switches and migration between cores in real-world applications as the cache behaviour is invariably affecting the performance.
Therefore, we use a simple microbenchmark to gauge the raw overhead.
Our microbenchmark executes a simple loop consisting solely of scalar instructions without any memory accesses.
For core specialization, 5\% of the loop is marked as if it was AVX code.
As the overhead depends on the frequency of migrations of tasks between cores, the length of the loop is varied.
The benchmark starts 26 threads and places them on 12 cores (24 hardware threads) of the system in order to achieve a similar environment compared to the web server benchmark.
As the remaining 4 cores are not used, we disable C-states to prevent the use of higher turbo boost frequencies which would cause the benchmark to underestimate overhead.

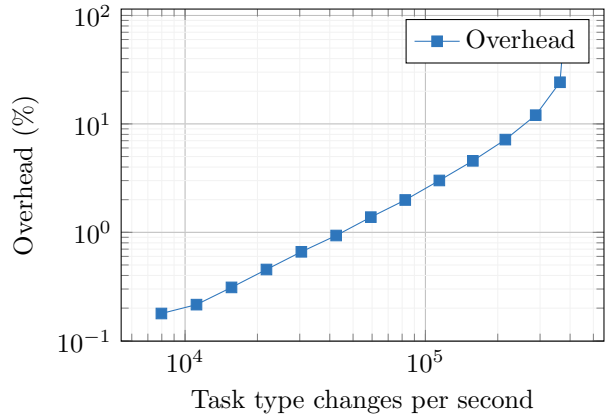
\begin{figure}
	\begin{tikzpicture}
		\begin{axis}[
			legend pos=north east,
			ylabel={Overhead (\%)},
			xlabel={Task type changes per second},
			width=8cm,
			height=6cm,
			grid style={line width=.1pt, draw=gray!10},
			major grid style={line width=.2pt,draw=gray!50},
			grid=both,
			ymode=log,
			xmode=log,
		]
		\addplot[color=kitblue,mark=square*] coordinates {

			(376399.87550513895, 63.64218767485062)
			(363098.26376065053, 24.21407868846815)
			(287747.28639230976, 12.031767577077488)
			(215066.76864398742, 7.163827800791282)
			(157539.00404501104, 4.565211414040249)
			(114274.98514448905, 3.0080256845795788)
			(82438.77514601732, 1.9875219299894127)
			(59269.55339900524, 1.3804813411159529)
			(42523.67067775152, 0.9346609593007713)
			(30466.39378756186, 0.6614037971453257)
			(21809.00606731722, 0.4541842838689689)
			(15601.708186689306, 0.3105710990237469)
			(11155.095625522701, 0.21538850042992408)
			(7970.798392481885, 0.17843300135942286)
		};
		\legend{Overhead}
		\end{axis}
	\end{tikzpicture}
	\caption{
		Overhead of core specialization in a CPU-intensive microbenchmark.
		The horizontal axis shows the task type changes (AVX vs. scalar) executed per second.
		For reference, the web server benchmark described above executes 55000 task type changes per second.
	}
	\label{fig:microbenchmark}
\end{figure}

Figure~\ref{fig:microbenchmark} plots the runtime overhead for different loop lengths.
The results show that over a wide range of task type change rates the overhead scales proportionally to the rate of task type changes.
The cost of each pair of switches from AVX to scalar code and back stays fairly constant (approximately 400-500\,ns), low enough to result in acceptable overhead for a wide range of applications.
Even at 100,000 task type changes per second (corresponding to 50,000 sections of AVX code), the overhead due to frequent scheduler invocations is below 3\%.

The result, however, also shows that in any situation the resulting performance impact depends on both frequency improvement as well as overhead.
Especially at higher task type change rates, the overhead can easily negate any positive effects.
Although our prototype always enables core specialization, we therefore expect that policies have to be adaptive to be viable for wide-spread use.
We expect that a good policy has to estimate the impact of core specialization on performance and, depending on the outcome, has to choose whether to use core specialization or not.

\section{Related Work}

We are, to the best of our knowledge, the first to implement core specialization to reduce the performance variation brought by wide vector instructions on current Intel CPUs.
However, we are not the first to describe the effects of these instructions, and core specialization has been successfully used in other scenarios to improve performance.
This section gives an overview over related work in both these areas.

\paragraph{AVX-Induced Frequency Changes}

Frequency variations depending on the executed instruction mix were first described for Haswell-EP processors which have different maximum frequencies depending on whether AVX instructions are executed\cite{hackenberg2015energy}.
Whereas previous CPUs operated at a constant frequency and the power consumption varied depending on the instruction mix, these CPUs keep their power consumption fairly constant, but frequency and therefore performance vary depending on the executed instructions.
In a cluster, this performance imbalance causes performance issues for tightly coupled code because significant amounts of time are spent in synchronization primitives such as barrier synchronization\cite{schuchart2016shift}.

The Skylake-SP microarchitecture and the introduction of the AVX-512 instruction set in server processors further increased the frequency variation, with frequency differences of up to 900\,MHz between AVX-512 code and scalar code\cite{xeonscalableerrata}.
As a result, in many cases the performance advantage of the wider vector registers when compared to AVX2 is negated by the frequency reduction.
For example, AVX-512 vectorization in the x265 video encoder did not yield any performance improvement for the \enquote{veryfast} quality profile when all cores were utilized, and even the \enquote{veryslow} profile only yielded a 10\% performance improvement even though the IPC gain was significantly higher\cite{tiwari2018accelerating}.
As future work, the authors therefore suggest either monitoring the CPU frequency to conditionally enable AVX-512 vectoring based on the CPU frequency or fundamental rearchitecting of the x265 encoder to let different cores execute different operations, with the intent to limit the number of cores which execute AVX-512 instructions.
We expect such a reengineering of the whole application to be costly and time-consuming, though.

Often, however, not the whole application is accelerated with wide vector instructions.
Instead, software engineers only apply vectorization where it seems fitting.
This situation can cause performance problems as the frequency reduction caused by AVX instructions is only reverted after at least approximately two milliseconds\cite{optimizationmanual}.
For example, web server benchmarks at Cloudflare showed significantly lower performance for a web server using AVX-512-enabled cryptography primitives in the OpenSSL library compared to their AVX2 counterparts in the BoringSSL library even though the OpenSSL library in isolation provided higher encryption and decryption speeds, simply because the frequency reduction affected unrelated web server code\cite{krasnovdangers}.

In a blog post, Daniel Lemire argues that a significant frequency reduction is only triggered by dense AVX code\cite{lemireavx}.
He suggest as future work that the operating system or an application framework could schedule threads on separate sets of cores whenever they execute sufficient numbers of AVX instructions..
However, no implementation or evaluation is described.
We provide an implementation of such a core specialization framework and provide an evaluation of its cost and benefits.
Especially we show that thread migration, when optimized for this use case, has low enough overhead to be a viable mechanism to limit the number of cores executing AVX-512 code.

\paragraph{Core Specialization}

We use core specialization as a technique to limit the effect of AVX-induced frequency reduction to select cores.
Core specialization has been suggested as a mechanism to increase performance before, although different effects were utilized.
As the fastest cache levels are usually private to the individual cores, core specialization can be used to place different parts of the system's working set in private caches of different cores, thereby increasing cache utilization by reducing the number of duplicated entries in different cores' private caches.
For example, FlexSC\cite{soares2010flexsc} places the operating system on a separate set of cores, whereas SchedTask\cite{kallurkar2017schedtask} analyzes the instruction footprint of code sections and uses instruction footprint similarity for scheduling decisions in order to reduce instruction cache misses.

Instead, our approach utilizes core specialization in order to limit the impact AVX-induced frequency changes to a subset of all cores.
We show that, although our prototype also shows a slight reduction of the amount of last-level cache misses due to core specialization, the impact on IPC is limited and the effects on processor frequency dominate system throughput.
However, it is likely that the approaches can be combined to further increase performance if the workload permits.

On single-core systems, an alternative to core specialization is to batch similar operations from multiple threads together and to schedule them so that the same part of the application code is executed repeatedly before other unrelated code pollutes the caches\cite{larus2002using,harizopoulos2004steps}.
If one codepath is repeatedly executed, starting from the second execution the instructions can be fetched from the cache which significantly reduces the number of instruction cache misses.
Similarly, batching should in theory reduce the frequency drop caused by AVX instructions if the AVX code from multiple threads is grouped together in order to reduce the number of frequency transitions.
However, even in this scenario, all cores periodically reduce their frequency.
Therefore, we expect this technique to have a lower performance benefit than our proposed mechanism.

\paragraph{OS Support for Heterogeneous Multiprocesors}

The approaches described above implement core specialization in software, but operate on multiprocessor systems containing hardware-wise identical cores.
However, different applications (or parts of applications) have different requirements to the underlying microarchitecture.
For example, a memory-intensive application might not be able to fully utilize the potential of a wide out-of-order architecture and might execute more efficiently on a more simple in-order system\cite{kumar2003single}.
As a result, single-ISA heterogeneous multi-core systems have been suggested which consist of cores with equal instruction sets but differing microarchitecture and operating frequency and provide an energy-efficient core for a wide range of applications\cite{kumar2003single}.
Especially heterogeneous applications with execution phases with significantly different behaviour can profit if the phases are each executed on their ideal core type.

Similarly, a heterogeneous multi-core system can provide cores with different ISAs\cite{venkat2014harnessing}.
For example, the ARM Thumb instruction set provides higher code density, but provides fewer and smaller general purpose registers, and is therefore efficient for execution of code sections with low register pressure, whereas core with higher register pressure executes more efficiently on architectures such as Alpha with larger register sets.
Also, the Thumb instruction set does not provide floating point and SIMD support which significantly improves peak power consumption and core area but requires costly emulation of floating point instructions.

Even though current server CPUs employ identical microarchitectures and instruction sets in all cores, we show that it is beneficial to artificially create heterogeneity even in these systems.
Limiting the execution of AVX2 and AVX-512 instruction to a subset of the cores lets all other cores execute at increased frequencies, thereby improving performance for code which does not use wide SIMD instructions.

On such a system, threads need to be migrated to a suitable core whenever they execute significant amounts of wide SIMD instructions.
\emph{Fault-and-migrate} is an operating system mechanism to automatically move threads to a suitable core\cite{li2010operating,gupta2013kinship}.
Whenever a thread executes an instruction not supported on its current core, the core triggers an undefined instruction exception.
Following the exception, the operating system selects a core with support for the instruction and migrates the thread.

Previous work assumes a heterogeneous multiprocessor where cores differ in hardware\cite{li2010operating,gupta2013kinship}.
However, the concept of fault-and-migrate is applicable to software-based heterogeneity similar to the one used in our design.
Whereas our prototype currently requires the developer to manually annotate code sections which make use of wide SIMD instructions, we intend to extend the prototype to make use of fault-and-migrate in order to automatically detect problematic code regions.
Li et al. describe how disabling the floating point unit allows to emulate instruction set asymmetry on current systems\cite{li2010operating}.
Similarly, we intend to restrict the size of the memory region used for the FXSTOR instruction during context switches\cite[Sec. 2.6.11]{intelmanualvol2} to selectively let AVX-512 instructions invoke the operating system.

Shen et al.\cite{shen2009novel} describe several scheduling algorithms for ISA-heterogeneous multiprocessors which share a common core ISA.
They describe a system in which each task has a \emph{task ISA ID} and where the scheduler prefers tasks whose task ISA ID closely matches the core's ISA.
We apply a similar approach to software-defined heterogeneity on current Intel server processors.
As our design is integrated into an existing scheduler with deadline-based priorities, we employ a slightly different priorization mechanism to prevent starvation of kernel tasks.

\section{Conclusion}\label{sec:conclusion}

Many recent server CPUs reduce their frequency when wide vector instructions are executed.
In order to reduce the number of frequency transitions, these CPUs delay reverting to the original frequency, so any scalar code following such vectorized portions of the code is significantly slowed down.
For heterogeneous workloads, this effect can reduce overall performance by over 10\%.

We propose core specialization to limit the impact of these frequency effects.
By limiting the subset of cores on which vectorized parts of the program are executed, only the frequency of those cores is affected.
We show that migration of unmodifed threads is a viable unobtrusive mechanism for core specialization, and we describe a scheduler interface which can be used to mark vectorized code regions so that the threads are then transparently migrated to a suitable core.
The evaluation of our prototype based on the MuQSS scheduler with workloads based on the nginx web server and the OpenSSL SSL library shows that core specialization can reduce the frequency impact of AVX instructions by over 70\%.

\subsection{Future Work}

Our prototype requires manual instrumentation of the program to mark vectorized parts of the code.
In this work, we describe how automatic disassembling of the program combined with a performance counter analysis can be used to identify these parts of the code.
We also describe a mechanism using the last-branch records found in recent CPUs to accurately identify very short sections of AVX-heavy code.
As our evaluation workload, however, did not contain any such short AVX code sections, we were not able to evaluate this approach.
We intend to evaluate a prototype of this technique with synthetic benchmarks to show its accuracy, and we intend to conduct a survey over a wider range of applications vectorized with AVX2 and AVX-512 to estimate the need for such techniques.

Although our prototype requires manual instrumentation by the user, ideally no such input would be required and the system would use fault-and-migrate techniques\cite{li2010operating} to automatically migrate AVX code to the appropriate cores.
We intend to evaluate restricting the size of the memory region used by the FXSTOR instruction to store the FPU content as a method to make wide instructions fault, at which point the threads can be migrated before any frequency reduction is triggered.

\bibliographystyle{abbrv}
\bibliography{techreport}

\begin{thebibliography}{10}

\bibitem{optimizationmanual}
{\em Intel® 64 and IA-32 Architectures Optimization Reference Manual}, Apr.
  2018.

\bibitem{intelmanualvol2}
{\em Intel® 64 and IA-32 Architectures Software Developer's Manual - Volume 2
  (2A, 2B, 2C \& 2D): Instruction Set Reference, A-Z}, May 2018.

\bibitem{xeonscalableerrata}
{\em Intel® Xeon® Processor Scalable Family -- Specification Update}.
\newblock Intel Corporation, Feb. 2018.

\bibitem{gregg2016flame}
B.~Gregg.
\newblock The flame graph.
\newblock {\em Communications of the ACM}, 59(6):48--57, 2016.

\bibitem{gupta2013kinship}
V.~Gupta, R.~Knauerhase, P.~Brett, and K.~Schwan.
\newblock Kinship: efficient resource management for performance and
  functionally asymmetric platforms.
\newblock In {\em Proceedings of the ACM International Conference on Computing
  Frontiers}, page~16. ACM, 2013.

\bibitem{hackenberg2015energy}
D.~Hackenberg, R.~Sch{\"o}ne, T.~Ilsche, D.~Molka, J.~Schuchart, and R.~Geyer.
\newblock An energy efficiency feature survey of the intel haswell processor.
\newblock In {\em Proceedings of the 2015 IEEE International Parallel and
  Distributed Processing Symposium Workshop}, pages 896--904. IEEE, 2015.

\bibitem{harizopoulos2004steps}
S.~Harizopoulos and A.~Ailamaki.
\newblock Steps towards cache-resident transaction processing.
\newblock In {\em Proceedings of the Thirtieth International Conference on Very
  Large Data Bases}, volume~30, pages 660--671. VLDB Endowment, 2004.

\bibitem{kallurkar2017schedtask}
P.~Kallurkar and S.~R. Sarangi.
\newblock Schedtask: a hardware-assisted task scheduler.
\newblock In {\em Proceedings of the 50th Annual IEEE/ACM International
  Symposium on Microarchitecture}, pages 612--624. ACM, 2017.

\bibitem{kleen2014lbr}
A.~Kleen.
\newblock An introduction to last branch records, Mar.~23, 2016.
\newblock \url{https://lwn.net/Articles/680985/}.

\bibitem{muqss}
C.~Kolivas.
\newblock Muqss - the multiple queue skiplist scheduler.
\newblock \url{http://ck.kolivas.org/patches/muqss/sched-MuQSS.txt}.

\bibitem{krasnovdangers}
V.~Krasnov.
\newblock On the dangers of intel's frequency scaling, Oct.~10, 2017.
\newblock
  \url{https://blog.cloudflare.com/on-the-dangers-of-intels-frequency-scaling/}.

\bibitem{kumar2003single}
R.~Kumar, K.~I. Farkas, N.~P. Jouppi, P.~Ranganathan, and D.~M. Tullsen.
\newblock Single-isa heterogeneous multi-core architectures: The potential for
  processor power reduction.
\newblock In {\em Proceedings of the 36th annual IEEE/ACM International
  Symposium on Microarchitecture}, page~81. IEEE Computer Society, 2003.

\bibitem{larus2002using}
J.~R. Larus and M.~Parkes.
\newblock Using cohort-scheduling to enhance server performance.
\newblock In {\em Proceedings of the USENIX 2002 Annual Technical Conference},
  pages 103--114. USENIX Association, 2002.

\bibitem{lemireavx}
D.~Lemire.
\newblock Avx-512: when and how to use these new instructions, Sept.~9, 2018.
\newblock
  \url{https://lemire.me/blog/2018/09/07/avx-512-when-and-how-to-use-these-new-instructions/}.

\bibitem{li2010operating}
T.~Li, P.~Brett, R.~Knauerhase, D.~Koufaty, D.~Reddy, and S.~Hahn.
\newblock Operating system support for overlapping-isa heterogeneous multi-core
  architectures.
\newblock In {\em 16th International Symposium on High Performance Computer
  Architecture}, pages 1--12. IEEE, 2010.

\bibitem{mazouz2014evaluation}
A.~Mazouz, A.~Laurent, B.~Pradelle, and W.~Jalby.
\newblock Evaluation of cpu frequency transition latency.
\newblock {\em Computer Science - Research and Development}, 29(3-4):187--195,
  2014.

\bibitem{molka2010characterizing}
D.~Molka, D.~Hackenberg, R.~Sch{\"o}ne, and M.~S. M{\"u}ller.
\newblock Characterizing the energy consumption of data transfers and
  arithmetic operations on x86-64 processors.
\newblock In {\em International Conference on Green Computing}, pages 123--133.
  IEEE, 2010.

\bibitem{schuchart2016shift}
J.~Schuchart, D.~Hackenberg, R.~Sch{\"o}ne, T.~Ilsche, R.~Nagappan, and M.~K.
  Patterson.
\newblock The shift from processor power consumption to performance variations:
  fundamental implications at scale.
\newblock {\em Computer Science - Research and Development}, 31(4):197--205,
  2016.

\bibitem{shen2009novel}
H.~Shen and F.~P{\'e}trot.
\newblock Novel task migration framework on configurable heterogeneous mpsoc
  platforms.
\newblock In {\em Proceedings of the 2009 Asia and South Pacific Design
  Automation Conference}, pages 733--738. IEEE Press, 2009.

\bibitem{soares2010flexsc}
L.~Soares and M.~Stumm.
\newblock {FlexSC}: Flexible system call scheduling with exception-less system
  calls.
\newblock In {\em Proceedings of the 9th USENIX Conference on Operating Systems
  Design and Implementation}, pages 33--46. USENIX Association, 2010.

\bibitem{taylor2012dark}
M.~B. Taylor.
\newblock Is dark silicon useful? harnessing the four horsemen of the coming
  dark silicon apocalypse.
\newblock In {\em 49th ACM/EDAC/IEEE Design Automation Conference}, pages
  1131--1136. IEEE, 2012.

\bibitem{tiwari2018accelerating}
P.~K. Tiwari, V.~V. Menon, J.~Murugan, J.~Chandrasekaran, G.~S. Akisetty,
  P.~Ramachandran, S.~K. Venkata, C.~A. Bird, and K.~Cone.
\newblock {Accelerating x265 with Intel\textregistered{} Advanced Vector
  Extensions 512}.
\newblock Technical report, Intel, 05 2018.

\bibitem{venkat2014harnessing}
A.~Venkat and D.~M. Tullsen.
\newblock Harnessing isa diversity: Design of a heterogeneous-isa chip
  multiprocessor.
\newblock {\em ACM SIGARCH Computer Architecture News}, 42(3):121--132, 2014.

\end{thebibliography}
\end{document}